\documentclass[aps,prd,amsfonts,amssymb,preprint,eqsecnum,nofootinbib,superscriptaddress]{revtex4}
\usepackage{amsmath, mathrsfs}
\usepackage{hyperref, graphicx, subfigure}

\newcommand\tr{\mathrm{Tr}\,}

\begin{document}

\title{
Pion condensation in holographic QCD
\vskip 0.1in}
\author{Dylan Albrecht}%\email[]{djalbrecht@email.wm.edu}
\author{Joshua Erlich}%\email[]{jxerli@wm.edu}

\affiliation{Particle Theory Group, Department of Physics,
College of William and Mary, Williamsburg, VA 23187-8795}
%
%\date{\today}

\newcommand\sect[1]{\emph{#1}---}

\begin{abstract}
We study pion condensation at zero temperature
in a hard-wall holographic model of hadrons with isospin chemical potential.
We find that the transition from the hadronic phase to the pion condensate 
phase is first order except in a certain limit of model parameters. 
Our analysis suggests that immediately across the phase boundary 
the condensate 
acts as a stiff medium approaching the Zel'dovich limit
of equal energy density and pressure.

\end{abstract}
\maketitle

\section{Introduction}

Dense nuclear matter generically carries net isospin and consequently has 
a nonvanishing isospin chemical potential, $\mu_I$.  
The Relativistic Heavy Ion Collider (RHIC) and the 
Large Hadron Collider (LHC) are presently 
the premier laboratories capable of studying the quark-gluon plasma phase of 
nuclear matter.  The heavy nuclei in these experiments
carry isospin, so the finite temperature 
system after collision has a 
nonvanishing isospin chemical potential.  Neutron stars have still
larger  chemical potential, and the behavior of such
objects depends on the phase structure of QCD in cold, dense environments
\cite{Baym:1973zk,Migdal:1979je,Haensel:1982zz,Steiner:2004fi}.
At large chemical potential it is expected 
that mesons condense, beginning with the pions at $\mu_I\sim m_\pi$ 
\cite{Migdal:1971cu,Migdal:1976zc,Sawyer:1973fv}.  
The phase diagram of
QCD at high density has been explored via the chiral Lagrangian
\cite{Son:2000xc,Splittorff:2002xn}, the Nambu-Jona-Lasinio model 
\cite{Toublan:2003tt,He:2005sp,Abuki:2008wm}, and
lattice QCD 
\cite{Kogut:2002zg,deForcrand:2007uz,Detmold:2008fn,Detmold:2008yn}.  
The general
consensus is that at low temperatures and vanishing baryon chemical 
potential there is a second order transition 
from the hadronic phase to a pion condensate phase  
at a critical isospin chemical potential $\mu_I$ around the
pion mass $m_\pi\approx 140$ MeV, or larger at finite temperature.

The AdS/CFT correspondence \cite{Maldacena:1997re,Witten:1998qj,Gubser:1998bc}
has provided motivation for 
extra-dimensional (holographic) models of QCD 
\cite{Polchinski:2000uf,Brodsky:2003px,deTeramond:2005su,Babington:2003vm,
Kruczenski:2003uq,Erlich:2005qh,Sakai:2004cn,Da Rold:2005zs,Hirn:2005nr}.
Both explicit and spontaneous chiral symmetry breaking may be built into 
the extra-dimensional 
models, resulting in an effective description similar to
extended hidden-local-symmetry models \cite{Bando:1985rf,Son:2003et}.  
The AdS/CFT correspondence
maps sources and expectation values of 
field theory operators to backgrounds of extra-dimensional fields with corresponding
quantum numbers.  By studying
fluctuations about a prescribed background, the model makes predictions 
for field theory observables.  

We study the behavior of matter with isospin
chemical potential in a holographic model of hadrons.  
Related analyses in other
holographic systems appear elsewhere ({\em e.g.} 
Refs.~\cite{Kim:2006gp,Aharony:2007uu,Basu:2008bh,Karch:2009eb}).  We
are motivated to study the hard-wall model of 
Refs.~\cite{Erlich:2005qh,Da Rold:2005zs}
in light of recent suggestions that the
pion condensate phase is absent in that model
\cite{Kim:2007gq}.
In contrast, we identify the pion condensate phase and study its properties.

The hard-wall models \cite{Polchinski:2000uf,Brodsky:2003px,
deTeramond:2005su,Erlich:2005qh,Da Rold:2005zs} are the
simplest holographic models which capture certain features of QCD.  
The hard-wall geometry is a region of 4+1 dimensional (5D)
Anti-de Sitter
space preserving the isometries of 3+1 dimensional (4D) Minkowski space.  
The
main motivation for this choice of spacetime is its simplicity, although
arguments have been made to support the choice of 
Anti-de Sitter space in light of asymptotic freedom at high energies
and  evidence for
an approximate conformal invariance in QCD at lower energies 
\cite{Brodsky:2008be,Deur:2008rf}.  The presence of a wall, which terminates
the spacetime at what is referred to as the IR boundary, leads to a
discrete spectrum of Kaluza-Klein modes 
identified via their quantum numbers with towers of hadronic resonances.

Global symmetries are lifted to gauge invariances in the holographic 
description.
In order to include the approximate chiral symmetry of the up and down quarks, 
SU(2)$_L\times$SU(2)$_R$
gauge fields are included in the 5D model.  To mimic the pattern of
chiral symmetry breaking, a set of scalar fields transforming in the
bifundamental representation of the chiral symmetry group is introduced.  
The quantum numbers of the scalar fields are those of the
scalar quark bilinear $\overline{q}_L^iq_R^j$, where $i$ and $j$
are flavor indices which are now gauge indices.  
The scalar fields have a background profile which
preserves the 4D Lorentz invariance of the spacetime but 
breaks the chiral symmetry to the diagonal isospin subgroup.

The linearized equations of motion for the scalar field have
two independent background solutions: a normalizable mode and a 
non-normalizable mode,
the difference being that the normalizable mode has a finite effective 4D 
action, while the non-normalizable mode has a divergent action.  If
the non-normalizable mode is turned on, then the theory is modified so as
to explicitly break the chiral symmetry, as if by a quark mass; 
if the normalizable mode has a nonvanishing background
then there is a spontaneous breaking of the chiral symmetry, as if by 
a contribution to the chiral condensate $\langle\overline{q}_L^iq_R^j\rangle$.

The chemical potential for the third component of isospin
acts as a source for the isospin number density \begin{equation}
N_3=q_L^{\dagger\,i}\, T^3_{ij}\,q_L^j+
q_R^{\dagger\,i}\, T^3_{ij}\,q_R^j,
\end{equation}
which is the time component of the isospin current
\begin{equation}
J_V^{a\,\nu}=\overline{q}^i\gamma^\nu T^a_{ij}q^j, \end{equation}
where $T^a=\sigma^a/2$ are the generators of the isospin SU(2), and
the subscript $V$ represents the vector subgroup of the chiral symmetry.
A source for the time component of the current 
couples as would the time component of a background gauge field.
Hence, a non-normalizable background for the time component of the
vector combination of 5D gauge fields mimics an isospin 
chemical potential in the 4D effective theory.

As the magnitude of the 
isospin chemical potential is increased above the pion mass, it becomes
energetically favorable for a pion condensate to form.  We find that 
the phase transition
is first order in the hard-wall model unless the  5D gauge coupling vanishes.
The speed of sound $c_s$ at high temperatures was conjectured to satisfy a
``sound bound''
$c_s^2<1/3$ \cite{Hohler:2009tv,Cherman:2009tw}, where $c_s^2=1/3$ is
the conformal limit.
Fluctuations in the condensate at zero temperature violate the ``sound bound,''
except near the phase boundary and then only if the 5D gauge coupling is small enough.  
Violation of the sound bound at low temperature is not unusual
\cite{Cherman:2009tw} and has also been observed in certain D-brane
systems \cite{Karch:2009eb} and in a holographic model describing matter
at a Lifshitz point \cite{Lee:2010uy}.

To describe systems at nonvanishing temperature, 
extra-dimensional models are modified to include
a black-hole horizon.  However, 
we will focus on the zero-temperature phase of isospin
matter, which corresponds to the original hard-wall background without a 
black-hole horizon.
For simplicity we do not include chemical potentials except for isospin, so
our analysis provides only a narrow cross section of the phase structure of
the model.  Extensions of these results to nonvanishing
temperature and baryon chemical potential,
and to include strange quarks and Kaon condensation
\cite{Kaplan:1986di}, may shed light on the phases
of matter in neutron stars and other extreme environments.

\section{Holographic Pion Condensation}

The action for the 5D hard-wall model with chiral symmetry is given by
\cite{Erlich:2005qh,Da Rold:2005zs},
\begin{equation}
\label{eaction}
S = \int d^{5}x \,\sqrt{-g} \, {\rm Tr} \left\{
	\left|DX\right|^{2} + 3\left| X \right|^{2} -
	\frac{1}{4 g_{5}^{2}} \left( F_{L}^{2} + F_{R}^{2} \right)
	\right\},
\end{equation}
where \( D_{M}X = \partial_{M}X - iL_{M}X + iXR_{M}\),
\(L_{M} = L_{M}^{a} T^{a}\) and 
\(F_{M N}^{L} = \partial_{M} L_{N} - \partial_{M} L_{N}
- i\left[L_{M}, L_{N} \right]\) (similarly for \(R\)), and we normalize the gauge 
kinetic term as in \cite{Erlich:2005qh}.
The spacetime in the hard-wall model is a slice of \(AdS_{5}\):
\begin{equation*}
ds = a(z)^{2} \left( \eta_{\mu \nu} dx^{\mu} dx^{\nu} - dz^{2} \right ),
	\qquad \epsilon <z\le z_{m},
\end{equation*}
where \(a(z) = 1/z\) in units of the AdS curvature scale,
and \(\eta_{\mu\nu}\) is the 4D Minkowski metric with mostly
negative signature.  Greek indices range from 0 to 3, and capital Latin
indices from 0 to 4, with $x^4$ also denoted by $z$.  The scalar fields $X$
transform in the bifundamental representation of the 
SU(2)$_L\times$SU(2)$_R$ gauge invariance.

Chiral symmetry breaking is provided by the background solution to the
\(X\) field equation of motion,
\begin{equation}
X_{0}(z) = \frac{1}{2} \left(m_{q} z + \sigma z^{3} \right) \equiv \frac{1}{2} v,
\end{equation}
where \(m_{q}\) is the quark mass matrix responsible for sourcing \(\sigma\),
the chiral condensate.  The bulk vector gauge field 
\(V_{M}^{a} = 1/2 (L_{M}^{a} + R_{M}^{a})\) is dual to the isospin vector 
current operator.
We work in the gauge $L_z^a=R_z^a=0$. 
The linearized equations of motion for the transverse part of
\(V_{\mu}^{a}\) are
\begin{equation}
\partial_{z} \left( \frac{1}{z} \partial_{z} V_{\mu}^{a}\right) - 
\frac{1}{z}\partial_\alpha\partial^\alpha V_\mu^a = 0 .
\end{equation}
The background solutions for $V_0^3$ are of the form \begin{equation}
V_{0}^{3}(z) = c_{1} + \frac{c_{2}}{2} z^{2},
\end{equation}
where the coefficient of the non-normalizable mode,
$c_1$, is identified with the chemical potential for the third component of
isospin $\mu_I$; and $c_2$ is 
proportional to the spontaneously generated background isospin number density,
which we assume to vanish.
Hence, the background gauge field is uniform, 
\begin{equation}
V_{0}^{3} = \mu_{I}.
\end{equation}

The pions are identified with solutions to the linearized coupled
equations of motion for the Goldstone modes in the scalar fields $X$, which mix
with the longitudinal part of the axial vector field 
\(A_\mu^a=(L_\mu^a-R_\mu^a)/2\equiv\partial_\mu\phi^a\).
We parametrize the Goldstone modes by fields $\pi^a$ such that,
\begin{equation}\begin{split}
X &= X_{0} \exp\left[i2\pi^{a} T^{a}\right] \\
 &= X_{0} \left(\cos b + i\left(n^{a} \sigma^{a} \right) \sin b\right),
\end{split}\label{xeqn}\end{equation}
where \(b = \sqrt{\pi^{c} \pi^{c}}\) and \(n^{c} = b^{-1} \pi^{c}\).  
The action \eqref{eaction} takes the form:
\begin{multline}
\label{newact}
S = \int d^{5} x \sqrt{-g} \biggl\{2 X_{0}^{2}\left( \partial_{M} (\cos b)
    \partial^{M} (\cos b)
  + \partial_{M} \left(n^{a} \sin b\right) \partial^{M} \left(n^{a}
    \sin b\right)
\right.
\biggr.
\\
  - 2 \mu_{I} a^{-2} \partial_{0} \left( n^{c} \sin b\right)
    \epsilon^{a3c} n^{a} \sin b
  - 2 a^{-2} \partial_{\mu} \left(n^{a} \sin b\right) \cos b\,
    \partial^{\mu}\phi^{a}
\\
  + 2 a^{-2} \partial_{\mu} (\cos b) n^{a} \sin b\, \partial^{\mu}\phi^{a}
  + 2 \mu_{I} a^{-2} \cos b\, \epsilon^{a3c} n^{a} \sin b\, \partial_{0}\phi^{c}
\\
\left.
  + \mu_{I}^{2} a^{-2} \sin^{2} b\, n^{c} n^{d} \epsilon^{c3e} \epsilon^{d3e}
  + a^{-2} \cos^{2} b\, \partial_{\mu}\phi^{a} \partial^{\mu}\phi^{a}
  + a^{-2} \sin^{2} b\, n^{a} \partial_{\mu}\phi^{a} n^{c} \partial^{\mu}\phi^{c}
    \right)
\\
\biggl.
  - a^{-4}\frac{1}{2 g_{5}^{2}}\left[ \mu_{I}^{2} \left(
    \partial_{i} \phi^{1} \partial^{i}\phi^{1} + \partial_{i} \phi^{2}
    \partial^{i} \phi^{2} \right)
  - \partial_{z} \partial_{\mu} \phi^{a} \partial_{z} \partial^{\mu} \phi^{a}
    +\mathcal{O}\left((A_\mu^a)^4\right)\right]
\biggr\},
\end{multline}
where contractions of Greek indices are done with \(\eta_{\mu \nu}\)
and those of capital Latin indices are done using the full metric \(g_{MN}\). 

To
quadratic order in \(\pi^{a}\) and \(\phi^{a}\), the action is
\begin{multline}
\label{quadaction}
S = \int d^{5} x \left\{ v^{2} a^{3} \left(\frac{1}{2} \partial_{\mu} \pi^{a}
    \partial^{\mu} \pi^{a} - \frac{1}{2} \partial_{z} \pi^{a} \partial_{z} \pi^{a}
  + \mu_{I} \left(\partial_{0} \pi^{2} \pi^{1} - \partial_{0} \pi^{1}
    \pi^{2} \right)
\right.
\right.
\\
\left.
  + \frac{1}{2} \mu_{I}^{2} \left(\pi^{1} \pi^{1} + \pi^{2} \pi^{2}\right)
  - \partial_{\mu} \pi^{a} \partial^{\mu} \phi^{a}
  + \frac{1}{2}\partial_{\mu} \phi^{a} \partial^{\mu} \phi^{a}
  + \mu_{I} \left(\pi^{2} \partial_{0} \phi^{1} - \pi^{1}
    \partial_{0}\phi^{2}\right)\right)
\\
\left.
  - a \frac{1}{2 g_{5}^{2}}\left[ \mu_{I}^{2} \left(
    \partial_{i} \phi^{1} \partial^{i}\phi^{1} + \partial_{i} \phi^{2}
    \partial^{i} \phi^{2} \right)
  - \partial_{z} \partial_{\mu} \phi^{a} \partial_{z} \partial^{\mu} \phi^{a}
  \right]
  \right\}.
\end{multline}
The linearized equations of motion for \(\pi^{a}\) and
\(\phi^{a}\) are:
\begin{equation}
\label{piphieom}
\begin{array} {l}
  \phi^{0} - \pi^{0} =
    \frac{1}{v^{2} a^{3} g_{5}^{2}} \partial_{z} \left(a \partial_{z}
    \phi^{0} \right),
\\
  m_{0}^{2} \phi^{0} - m_{0}^{2} \pi^{0}
    = \frac{1}{v^{2} a^{3}}
    \partial_{z} \left(v^{2} a^{3} \partial_{z} \pi^{0} \right),
\\ \quad \\
  m_{\pm} \phi^{\pm} - \left(m_{\pm} \mp \mu_{I}\right) \pi^{\pm} =
    \frac{m_{\pm}}{v^{2} a^{3} g_{5}^{2}} \partial_{z} \left(a \partial_{z}
    \phi^{\pm} \right),
\\
  \left(m_{\pm}^{2} \mp \mu_{I} m_{\pm} \right) \phi^{\pm}
    -\left(m_{\pm}^{2} \mp 2 \mu_{I} m_{\pm} + \mu_{I}^{2} \right) \pi^{\pm}
    = \frac{1}{v^{2} a^{3}}
    \partial_{z} \left(v^{2} a^{3} \partial_{z} \pi^{\pm} \right),
\end{array}
\end{equation}
where \(\pi^{1} = \frac{1}{\sqrt{2}} (\pi^{+} + \pi^{-})\),
\(\pi^{2} = \frac{-i}{\sqrt{2}} (\pi^{+} - \pi^{-})\), and \(\pi^{3} = \pi^{0}\)
(similarly for \(\phi^{a}\)).  These equations are evaluated in the pion rest
frame (\(\vec{q} = 0\)), identifying the pion frequency
with the effective pion mass in the isospin background.
Working in the gauge $A_{L\,z}^a=A_{R\,z}^a=0$, the 
fields satisfy the boundary conditions
\(\partial_{z} \phi^{\pm,0}(z_{m}) = \phi^{\pm,0}(\epsilon) =
\pi^{\pm,0}(\epsilon) = 0\).  The boundary condition at $z_m$ corresponds to
the gauge-invariant condition $F^L_{z\mu}(z_m)=F^R_{z\mu}(z_m)=0$, 
but this choice is not unique and is made for simplicity.

By eliminating \(\phi^{\pm, 0}\) we obtain equations of
motion for $\pi^{\pm,0}$ alone.
\begin{equation}
\label{pieom}
\begin{array} {l}
\partial_{z} \left( \frac{1}{v^{2}a^{3}} \partial_{z}
	  \left(v^{2}a^{3} \partial_{z} \pi^{0} \right) \right)
	+ m_{0}^{2} \partial_{z}\pi^{0}
	- g_{5}^{2} v^{2} a^{2} \partial_{z}
	  \pi^{0} = 0,
  \\
\partial_{z} \left( \frac{1}{v^{2}a^{3}} \partial_{z}
	  \left(v^{2}a^{3} \partial_{z} \pi^{\pm} \right) \right)
	+ \left( m_{\pm}^{2} \mp 2 \mu_{I} m_{\pm} + \mu_{I}^{2}\right) 
	  \partial_{z}\pi^{\pm}
  \\
    \qquad \qquad
	- g_{5}^{2} v^{2} a^{2} \partial_{z}
	  \pi^{\pm} = 0.
\end{array}
\end{equation}

Except for the replacement of the eigenvalue $m_0^2$ with  
$\left( m_{\pm}^{2} \mp 2 \mu_{I} m_{\pm} + \mu_{I}^{2}\right)$, the
fields \(\pi^{\pm}\) and $\pi^0$ 
are solutions to the same differential
equation with the same boundary conditions, with identical eigenvalues 
\begin{equation}m_0^2=(m_{\pm}^2\mp2\mu_Im_\pm+\mu_I^2).\end{equation}  
The neutral pion is unaffected by the chemical potential, so 
identifying $m_0$ with the pion mass in vacuum, $m_\pi$,
we find a relation for the masses of the charged pions:
\begin{equation}
m_{\pm} = \pm \mu_{I} + m_{\pi},
\end{equation}
where \(m_{\pi} \equiv m_{0}\).
For $|\mu_{I}|>m_\pi$ a charged pion mass
becomes negative, indicating the instability to pion condensation.  

\section{Properties of the pion condensate phase}
Since the pattern of chiral symmetry breaking
is built into our holographic model, we expect to reproduce predictions
of the chiral Lagrangian, at least qualitatively.
The pion effective theory is determined by the action on
the solution for the pion mode discussed in the previous section, integrated
over the extra dimension.

\subsection{Decoupling the 5D gauge fields}
The limit $g_5\rightarrow0$ provides the most direct comparison to previous
results.  In that limit the fluctuations of the 5D gauge fields decouple
from the pion physics.  The corresponding 4D effective theory is similar to the 
chiral Lagrangian with isospin chemical potential included
as a background for a 4D  isospin gauge field, as in Ref.~\cite{Son:2000xc}.
In terms of the unitary fields \begin{equation}
\Sigma=\exp\left[\frac{i\pi^a \sigma^a}{f_\pi}\right], \end{equation}
the leading order chiral Lagrangian is
\begin{equation}
\label{4DSS}
\mathcal{L}_{4D} = \frac{f_{\pi}^{2}}{4} {\rm Tr}\left(\nabla_{\nu} \Sigma
   \, \nabla^{\nu} \Sigma^{\dagger} \right) + \frac{m_{\pi}^{2} f_{\pi}^{2}}{4}
    {\rm Tr}\left(\Sigma + \Sigma^{\dagger}\right),
\end{equation}
where \(\nabla_{0}\Sigma = \partial_{0} \Sigma - i \frac{\mu_{I}}{2}
\left[\sigma_{3}, \Sigma\right]\) and \(\nabla_{i} = \partial_{i}\).  
Expanding to second order in the pion fields, the Lagrangian takes the form,
\begin{equation}
\begin{array} {l}
  \mathcal{L}_{4D} =
    \frac{1}{2} \partial_{\mu} \pi^{a} \partial^{\mu} \pi^{a}
    - \frac{1}{2}\left(m_{\pi}^{2} - \mu_{I}^{2}\right)
    \left(\pi^{1} \pi^{1} + \pi^{2} \pi^{2}\right)
  \\ \qquad
    - \frac{1}{2} m_{\pi}^{2} \pi^{3} \pi^{3}
    + \mu_{I} \left(\partial_{t}\pi^{1} \pi^{2}
    - \partial_{t}\pi^{2} \pi^{1} \right).
\label{eq:L2}\end{array}
\end{equation}
The instability when $|\mu_I|>m_\pi$ signals the phase transition to a pion 
condensate phase.  Estimation of
the value of the condensate and related observables 
requires an extension of the analysis to higher
order in the pion fields, which we perform in the holographic description.

By design, the analysis of the 5D model is similar to the chiral Lagrangian analysis
above.  
In the limit $g_5\rightarrow0$, we neglect couplings to the longitudinal
gauge field $\partial_\mu\phi^a$.  The action \eqref{newact} takes the
form
\begin{multline}
\label{eq:newact2}
S_{g_5=0} = \int d^{5} x \sqrt{-g} \biggl\{2 X_{0}^{2}\left( \partial_{M} (\cos b)
    \partial^{M} (\cos b)
  + \partial_{M} \left(n^{a} \sin b\right) \partial^{M} \left(n^{a}
    \sin b\right)\right)
\\
  - 2 \mu_{I} a^{-2} \partial_{0} \left( n^{c} \sin b\right)
    \epsilon^{a3c} n^{a} \sin b
  + \mu_{I}^{2} a^{-2} \sin^{2} b\, n^{c} n^{d} \epsilon^{c3e} \epsilon^{d3e}
   \biggr\},
\end{multline}
where $b=\sqrt{\pi^c\pi^c}$ and $n^c=b^{-1}\pi^c$ as before.  
The linearized equations of motion for the pion fields are now,
\begin{equation}
\label{eq:pi-nog5}
  - m_{\pi}^{2} \pi^{0,\pm}
    = \frac{1}{v^{2} a^{3}}
    \partial_{z} \left(v^{2} a^{3} \partial_{z} \pi^{0,\pm} \right).
\end{equation}
The condensate
is a static configuration rotationally invariant in $x^1,x^2,x^3$.  The action
on such configurations gives the condensate 
effective potential, 
\begin{multline}
V_{eff,g_5=0}=\int dz\,v(z)^2a(z)^3
\biggl(\frac{1}{2}\left(\frac{db}{dz}\right)^2+
\frac{1}{2}\sin^2b\left(\frac{dn^c}{dz}\right)^2
-\frac{\mu_I^2}{2}
\sin^2b \,n^c n^d\left(\delta^{cd}-\delta^{c3}\delta^{d3}\right)
\biggr). \label{eq:Veff-nog5}\end{multline}
The effective potential increases with $|dn^c/dz|$, 
so $dn^c/dz=0$ in the ground state.  The profile of $b(z)$ is determined
from the solution to the equations of motion for the pion Kaluza-Klein mode.
Expanding to fourth order in the pion fields,
\begin{multline}
V_{eff,g_5=0}=
\int_\epsilon^{z_m} dz\,v^2 /z^3\,\frac{1}{2}\left(\left(\frac{d\pi}{dz}\right)^2-
\mu_I^2 \left(\pi(z)^2-\frac{\pi(z)^4}{3}+\cdots\right)\,n^c n^d
\left(\delta^{cd}-\delta^{c3}\delta^{d3}\right)\right) \\
=\int_\epsilon^{z_m} dz\,v^2 /z^3\,\frac{1}{2}\left(m_\pi^2 \pi(z)^2-
\mu_I^2\,n^c n^d \left(\delta^{cd}-\delta^{c3}\delta^{d3}\right)
\left(\pi(z)^2-\frac{\pi(z)^4}{3}+\cdots\right)\right), \end{multline}
where we used the linearized equation of motion \eqref{eq:pi-nog5}
in the last line.

For $|\mu_I|>m_\pi$ it is energetically favorable to turn on the charged pions.
The pion field is normalized by its kinetic term in the effective 4D theory,
so we define $\pi^a(z)=\tilde{\pi}(z)\,\pi^a$ such that \begin{equation}
\int_\epsilon^{z_m} dz\,v^2 a^3 \,\tilde{\pi}(z)^2=1, \label{eq:pinorm}\end{equation}
and $\pi^a$ is the pion condensate 
$\langle\pi^a\rangle$.

Minimizing $V_{eff}$ expanded to $\mathcal{O}\left((\pi^a)^4\right)$, we find that 
the transition is smooth (second order), 
and for $\mu\gtrsim m_\pi$ we obtain, \begin{equation}
\pi^+\pi^-=\frac{3}{4\tilde{\eta}}\left(1-\frac{m_\pi^2}{\mu^2}\right),
\end{equation}
where $\tilde{\eta}=\int_\epsilon^{z_m}
 dz\,v^2 a^3 \tilde{\pi}(z)^4$.  We then find, \begin{equation}
V_{eff,g_5=0}(\pi^\pm)=-\frac{3}{8\tilde{\eta}}
\mu_I^2\left(1-\frac{m_\pi^2}{\mu_I^2}\right)^2. \end{equation}
The isospin number density is 
\begin{equation}
n_I=-\frac{\partial V_{eff}}{\partial\mu_I}=\frac{3\mu_I}{4\tilde{\eta}}
\left(1-
\frac{m_\pi^4}{\mu_I^4}\right). \end{equation}

We can express $\tilde{\eta}$ in terms of $f_\pi^2$ in this model by the AdS/CFT
determination of $f_\pi$.  The correlator of a product of axial currents
has a pion pole at zero momentum transfer in the chiral limit, 
with residue equal to $f_\pi^2$.
The AdS/CFT correspondence determines the correlation function in terms
of a bulk-to-boundary propagator which solves the linearized
equations of motion for
the transverse part of the axial vector field.  For more details in the 
context of the present model, see Refs.~\cite{Erlich:2005qh,Da Rold:2005zs}.  We
summarize the results here.

The linearized
equation of motion for the transverse part of the axial vector field
\(A_{\mu}^{a}(q,z)\) is given by,
\begin{equation}
\label{aperp}
\left[\partial_{z} \left(a \partial_{z} A_{\mu}^{a}\right)
  + \frac{q^{2}}{z} A_{\mu}^{a}
  - v^{2} a^{3} g_{5}^{2} A_{\mu}^{a}\right]_{\perp} = 0.
\end{equation}
The bulk-to-boundary propagator $A(q,z)$ describes the solution
to \eqref{aperp} of the form
\(A_{\mu}^{a}(q, z) = A(q, z) A_{0 \mu}^{a}(q)\), with boundary conditions
$\partial_z A(q,z)|_{z_m}=0$ and \(A(q, \epsilon) = 1\).
In terms of the bulk-to-boundary propagator the AdS/CFT prediction for the
pion decay constant is \begin{equation}
  f_{\pi}^{2} = -\frac{1}{g_{5}^{2}} \left. \frac{\partial_{z} A(0,z)}{z}
    \right|_{z=\epsilon}.
\label{eq:fpiAdSCFT}\end{equation}

\begin{figure}[ht]
  \centering
  \scalebox{.4}{\includegraphics{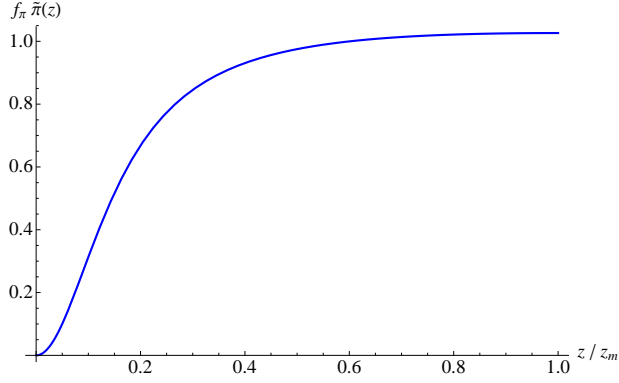}}
  \caption{Pion eigenfunction with \(m_{q} = 4.25\ {\rm MeV}\),
    \(\sigma = (263\ {\rm MeV})^{3}\), and $z_m=1/(323\ {\rm Mev})$.}
  \label{fig:piofz}
\end{figure}

If $g_5=0$ then the bulk-to-boundary propagator at $q^2=0$ is uniform, $A(0,z)=1$.  
To next order in $g_5^2$, we obtain
\begin{equation}
\frac{1}{z}\partial_z A(0,z)=
-g_5^2\int_z^{z_m} d\tilde{z}\,v(\tilde{z})^2/\tilde{z}^3+\mathcal{O}(g_5^4/z_m^4).
\end{equation}
From \eqref{eq:fpiAdSCFT} we obtain in the $g_5\rightarrow 0$ limit,
\begin{equation}\begin{array}{l}
f_\pi^2=\int_\epsilon^{z_m}dz\,v(z)^2/z^3 \\
\quad\approx\frac{\sigma^2z_m^4}{4}+m_q\sigma z_m^2+m_q^2\,\log(z_m/\epsilon).
\label{eq:fpi-nog5}\end{array}\end{equation}
Note that in the absence of boundary counterterms we must choose $\epsilon$ such
that $\log(z_m/\epsilon)\ll\sigma z_m^2/m_q$ to respect the chiral limit. We choose
 $\epsilon=1/(10^8\ {\rm MeV})$.  In that case
the integral defining $f_\pi^2$ is dominated by the region where the pion wavefunction
$\tilde{\pi}(z)$ is approximately constant.  Comparing \eqref{eq:fpi-nog5} with
\eqref{eq:pinorm}, we learn that the pion wavefunction $\widetilde{\pi}(z)\approx
1/f_\pi$ except for a region of small $z$, as in Figure \ref{fig:piofz}. 
Similarly, from the integral defining
$\tilde{\eta}$ we have \begin{equation}
\tilde{\eta}\approx 1/f_\pi^2.\end{equation}
For a concrete example, fixing the mass of the lightest KK mode of the vector field 
$V_\mu^a$ to 776 MeV determines 
$z_m=1/(323\ {\rm MeV})$ \cite{Erlich:2005qh}.  Then with $m_q$=4.25 MeV and
chiral condensate $\sigma=(263\ {\rm MeV})^3$ we find physical values 
$m_\pi$=140 MeV and $f_\pi$=92 MeV in the $g_5\rightarrow0$ limit.  With these
values of the parameters, we find \begin{equation}
\tilde{\eta}=\int_\epsilon^{z_m} dz\,v(z)^2 a(z)^3 \tilde{\pi}(z)^4 =1/(91\ {\rm MeV})^2,
\end{equation}
which is approximately $1/f_\pi^2$ as expected.  Note that the 
Gell-Mann-Oakes-Renner relation is approximately satisfied,
$m_\pi^2 f_\pi^2/(2m_q\sigma)=1.07\approx 1$.
 
\begin{figure}[ht]
  \centering
  \scalebox{.5}{\includegraphics{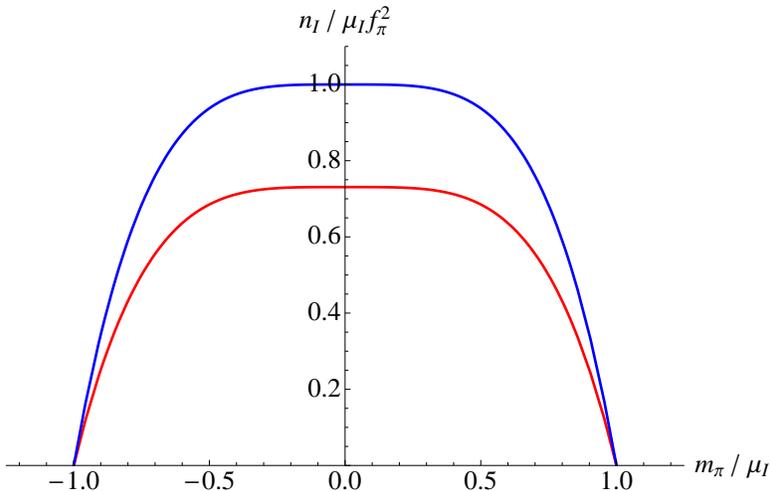}}
  \caption{Isospin number density. The bottom red curve is the prediction
of the hard-wall
    model with \(m_{q} = 4.25\ {\rm MeV}\), \(\sigma = (263\ {\rm MeV})^{3}\)
    and $z_m=1/(323\ {\rm MeV})$.
The top blue curve is the result from Ref. \cite{Son:2000xc} quoted in
\eqref{eq:n4D}.}
\label{fig:numberdensityglimit}
\end{figure}

We now have the holographic prediction of the equation of state:
\begin{equation}
n_I\approx\frac{3}{4}f_\pi^2\mu_I
\left(1-
\frac{m_\pi^4}{\mu_I^4}\right). \label{eq:nI-nog5}\end{equation}
For comparison, the corresponding prediction 
from the 4D chiral Lagrangian \eqref{4DSS} 
is \cite{Son:2000xc} \begin{equation}\label{eq:n4D}
n_{4D}=f_\pi^2 \mu_I\left(1-\frac{m_\pi^4}{\mu_I^4}\right), \end{equation}
which differs from the holographic prediction by an overall factor of $4/3$.
This overall factor drops out of the ratio of pressure to energy density
and the speed of sound  at zero temperature.  The number densities are
plotted in Figure \ref{fig:numberdensityglimit}.  The model is not expected
to be valid for $\mu_I\gtrsim m_\rho\approx 5.5 m_\pi$, 
but we plot the model prediction here and
below over the entire range of $\mu_I$.

The pressure $p$ and energy density $\varepsilon$ in the pion condensate medium
are determined by $n_I(\mu_{I})$ \cite{Son:2000xc,Kogut:2002zg}:
\begin{equation}
p(\mu_{I})
=\int_{m_\pi}^{\mu_{I}} n_{I}\,d\tilde{\mu}=\frac{3f_\pi^2\left(\mu_{I}^2
-m_\pi^2\right)^2}{8\mu_{I}^2},
\end{equation}
\begin{equation}
\varepsilon(\mu_{I})=\int_0^{n_{I}}\mu_{I}\,d\tilde{n}=\frac{3f_\pi^2}{8\mu_{I}^2}\left(\mu_{I}^2-m_\pi^2\right)
\left(\mu_{I}^2+3m_\pi^2\right). \end{equation}
This gives \begin{equation}
\frac{p}{\varepsilon}=\frac{\mu_{I}^2-m_\pi^2}{\mu_{I}^2+3m_\pi^2},
\end{equation}
and \begin{equation}
c_s^2=\frac{dp}{d\varepsilon}=
\frac{\mu_{I}^{4} - m_\pi^4}{\mu_{I}^{4} + 3 m_\pi^4}.\end{equation}
The speed of sound violates the sound bound $c_s^2<1/3$ except near the phase
transition boundary at $\mu_{I}=m_\pi$.

\subsection{Pion condensation with dynamical 5D gauge bosons}
Having understood how the $g_5\rightarrow0$ limit of holographic QCD
reproduces the
qualitative behavior of isospin matter at low temperature 
expected from the chiral Lagrangian, we 
now consider the more general situation including couplings to the 5D gauge fields.
In the calculations below we will take $g_5=2\pi$, which makes the 
holographic prediction of the vector current polarization at large momentum
transfer agree with perturbative three-color QCD \cite{Erlich:2005qh,Da Rold:2005zs} .

We first construct an approximate solution
to \eqref{piphieom} as an expansion in \(m_{\pi}\), as in Ref.~\cite{Erlich:2005qh}.  
Combining the first
two equations of \eqref{piphieom} we get
\begin{equation}
m_{\pi}^{2} \partial_{z} \phi^{0}
  = v^{2} a^{2} g_{5}^{2} \partial_{z} \pi^{0}.
\end{equation}
Recalling the boundary conditions $\pi(\epsilon)=\phi(\epsilon)=\partial_z\phi|_{z_m}=0$,
to zeroth order in $m_\pi$ the
solution is \(\pi^{0} = 0\), and \(\phi^{0}\) satisfies the
same equation as the bulk-to-boundary propagator in \eqref{aperp}, so we set
\(\phi^{0}(z) = A(0, z) - 1\).  Away from the boundary at $z=\epsilon$
consistency of the approximate solution with the first of the equations 
in \eqref{piphieom} requires
\(\pi^{0} \approx -1\).  
Recalling that the charged pions have the same wavefunction
as the neutral pion, we temporarily normalize the fields so that 
\(\pi^{\pm}(z) = \pi^{0}(z)\).  Then
\(\phi^{\pm} = (1 \mp \mu_{I} / m_{\pm}) \left[A(0, z) - 1\right]\).

Integrating the action \eqref{quadaction} by parts we get,
\begin{multline}
S = \int d^4 x dz \left\{\left[v^2 a^3 \left(\pi^{+} \pi^{-}
    - \pi^{+} \phi^{-} - \pi^{-} \phi^{+}
    + \phi^{+} \phi^{-}\right) - \frac{1}{g_{5}^{2}} \partial_{z} (a \partial_{z}
    \phi^{-})\phi^{+}\right] \partial_{\mu} \pi^{+}(x)
    \partial^{\mu} \pi^{-}(x)
\right.
\\
  + \left[\partial_{z}
    \left(v^{2} a^{3} \partial_{z} \pi^{+}\right) \pi^{-}
  + \mu_{I}^{2} v^{2} a^{3} \pi^{+} \pi^{-}\right] \pi^{+}(x) \pi^{-}(x)
\\
\left.
  -i \left[v^{2} a^{3} \mu_{I} \left(\pi^{-} \pi^{+} - \phi^{-} \pi^{+} \right)
    \right]\partial_{t} \pi^{-}(x) \pi^{+}(x)
  +i \left[v^{2} a^{3} \mu_{I} \left(\pi^{+} \pi^{-} - \phi^{+} \pi^{-} \right)
    \right]\partial_{t} \pi^{+}(x) \pi^{-}(x)
  \right\},
\end{multline}
where functions without an argument are understood to be functions of z.
Ignoring the \(\pi^{0}\) terms, which can be obtained by taking
\(\mu_{I} \rightarrow 0\), we can use the third of Eqs.~\eqref{piphieom} 
to solve for \(\phi^{\pm}\) and obtain
\begin{multline}
S = \int d^{4}x \left\{\left[-\frac{\mu_{I}^{2}}{m_{-} m_{+}} 
    \int dz \, v^{2} a^{3} \pi \pi
  - \frac{1}{g_{5}^{2}} \int dz\, \partial_{z}(a \partial_{z} \phi^{-}) \pi
\right.
\right.
\\
\left.
\left.
  + \frac{\mu_{I}}{m_{-}} \frac{1}{g_{5}^{2}} \int dz \,
    \partial_{z}(a \partial_{z} \phi^{+}) \pi \right] \partial_{\mu} \pi^{+}(x)
    \partial^{\mu} \pi^{-}(x)
\right.
\\
  + \left[\frac{m_{+}}{g_{5}^{2}} \int dz \partial_{z}(a \partial_{z} \phi^{+})
    \pi + \mu_{I}^{2} \int dz \, v^{2} a^{3} \pi \pi \right] \pi^{+}(x) \pi^{-}(x)
\\
  + i \left[\frac{\mu_{I}}{m_{-}} \int dz \, v^{2} a^{3} \pi \pi
    + \frac{1}{g_{5}^{2}} \int dz \, \partial_{z}(a\partial_{z} \phi^{-}
    \pi) \right] \partial_{t} \pi^{-}(x) \pi^{+}(x)
\\
\left.
  - i \left[-\frac{\mu_{I}}{m_{+}} \int dz \, v^{2} a^{3} \pi \pi
    + \frac{1}{g_{5}^{2}} \int dz \, \partial_{z} (a \partial_{z} \phi^{+})
    \pi\right] \partial_{t} \pi^{+}(x) \pi^{-}(x)
\right\},
\end{multline}
where we have used \(\pi \equiv \pi^{0}(z) = \pi^{\pm}(z)\).
We now use the approximate solutions for \(\phi^{\pm}\) and make the
approximation \(\pi = -1\) in those integrals dominated by the region where
the pion wavefunction is flat.  Defining
\(\alpha \equiv \frac{1}{f_{\pi}^{2}} \int dz \, v^{2} a^{3} \, \pi \pi\) and
making use of equation \eqref{eq:fpiAdSCFT}, we get
\begin{multline}
S = \int d^4 x \left\{
\left(\frac{m_{\pi}^{2} - \alpha \mu_{I}^{2}} {m_{\pi}^{2} - \mu_{I}^{2}} \right)
    f_{\pi}^{2} \partial_{\mu} \pi^{+}(x) \partial^{\mu} \pi^{-}(x)
  - \left(m_{\pi}^{2} - \alpha \mu_{I}^{2}\right) f_{\pi}^{2} \pi^{+}(x)
    \pi^{-}(x)
\right.
\\
\left.
  - 2 i \mu_{I} f_{\pi}^{2} \left(\frac{m_{\pi}^{2} - \alpha \mu_{I}^{2}}
    {m_{\pi}^{2} - \mu_{I}^{2}}\right) \partial_{t} \pi^{+}(x) \pi^{-}(x)
  \right\}.
\end{multline}
The resulting Lagrangian after canonically normalizing the kinetic term is,
transforming back to \((\pi^{1}, \pi^{2}, \pi^{3})\),
\begin{equation}
\begin{array} {l}
  \mathcal{L}_{eff} =
    \frac{1}{2} \partial_{\mu} \pi^{a} \partial^{\mu} \pi^{a}
    - \frac{1}{2}\left(m_{\pi}^{2} - \mu_{I}^{2}\right)
    \left(\pi^{1} \pi^{1} + \pi^{2} \pi^{2}\right)
  \\ \qquad
    - \frac{1}{2} m_{\pi}^{2} \pi^{3} \pi^{3}
    + \mu_{I} \left(\partial_{t}\pi^{1} \pi^{2}
    - \partial_{t}\pi^{2} \pi^{1} \right).
\end{array}
\end{equation}
This agrees with the leading order
4D chiral Lagrangian \eqref{eq:L2}.  

The effective potential for static configurations of $\pi(x)$
takes the same form as \eqref{eq:Veff-nog5} because the longitudinal
gauge bosons are derivatively coupled.
We make the ansatz that \(n^{3} = 0\); the
pion expectation value is only in the \((\pi^{1}, \pi^{2})\) plane.  
With the approximate solutions for $\pi(z)$ and $\phi(z)$ described above,
and applying the canonical rescaling of the kinetic term, the effective
potential at $\mathcal{O}(b^4)$ becomes
\begin{equation}
  V_{eff}\left(b\right) = \frac{1}{2} \left(m_{\pi}^{2} -
    \mu_{I}^{2}\right) b^{2} +
    \frac{1}{6} \mu_{I}^{2} f_{\pi}^{-2} \left(
    \frac{m_{\pi}^{2} - \mu_{I}^{2}}{m_{\pi}^{2} - \alpha \mu_{I}^{2}}
    \right)^{2} \eta \, b^{4},
\end{equation}
where \(\eta = \frac{1}{f_{\pi}^{2}} \int dz \, v^{2} a^{3} \,
\pi \pi \pi \pi\), \(\alpha = \frac{1}{f_{\pi}^{2}} \int dz \,
v^{2} a^{3} \, \pi \pi\), and \(f_{\pi}\) is given by
\eqref{eq:fpiAdSCFT}.  Note that $\pi(z)$ here is the approximate
solution for small $m_\pi$ described earlier, 
and is not the canonically normalized field.  
We can now solve for the 
minimum of the potential, \(\partial V_{eff} / \partial b = 0\):
\begin{equation}
b_{0}^{2} = \frac {3}{2} \frac{f_{\pi}^{2}}{\eta \mu_{I}^{2}}
    \frac{\left(m_{\pi}^{2} - \alpha \mu_{I}^{2}\right)^{2}}
    {\left(\mu_{I}^{2} - m_{\pi}^{2}\right)}.
\end{equation}
The isospin number density is
\begin{equation}
  \label{numberdensity}
  n_{I} = -\frac{\partial V_{eff}}{\partial \mu_{I}} =
    \mu_{I} f_{\pi}^{2} \frac{1}{\eta} \frac{3}{4} \left(\alpha^{2}
    - \frac{m_{\pi}^{4}}{\mu_{I}^{4}}\right).
\end{equation}
\begin{figure}[ht]
  \centering
  \scalebox{.5}{\includegraphics{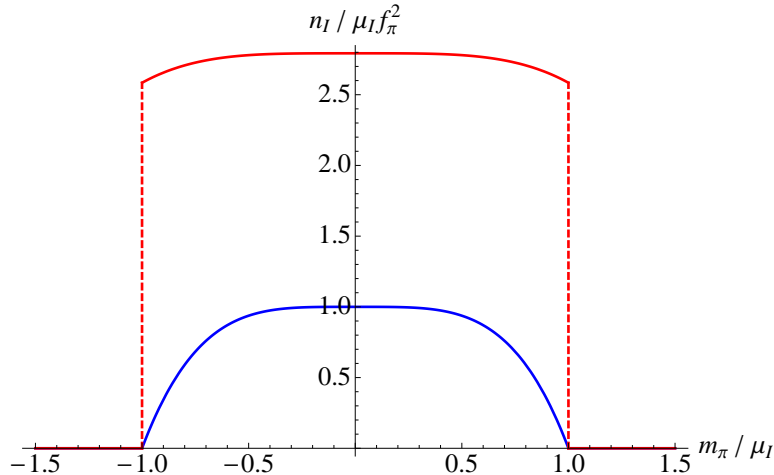}}
  \caption{Isospin number density. The top red curve is the perturbative 
prediction of the hard-wall
    model with parameters fit to $m_\rho$, $m_\pi$ and $f_\pi$:
\(m_{q} = 2.26\ {\rm MeV}\), \(\sigma = (333\ {\rm MeV})^{3}\),
 and $z_m=1/(323\ {\rm MeV})$.
This set of parameters gives \(\alpha = 3.66\) and
    \(\eta = 3.60\), in \eqref{numberdensity}. The bottom blue curve is the
    result from Ref. \cite{Son:2000xc} given in \eqref{eq:n4D}.}
 \label{fig:numberdensity}
\end{figure}
The phase transition is first order, with the order parameter jumping at
the phase boundary $\mu_I=m_\pi$.  
The minimum value of the free energy $V_{eff}$ 
at $\mathcal{O}(b^4)$ is discontinuous at the critical point
$\mu_I=m_\pi$, but the perturbative expansion breaks down near 
the critical point 
and we expect a nonperturbative analysis to confirm continuity in the free 
energy across the phase boundary.  
However, we can say with confidence that the transition
is not second order in this model, because if it were then the order parameter 
$b$ would vary smoothly and we would expect a perturbative analysis to 
be valid.  Although perturbation theory breaks down  near the
transition, we plot the perturbative prediction for the number density
in Figure \ref{fig:numberdensity}.

The ratio of the pressure to energy density is now
\begin{equation}
  \frac{p}{\varepsilon} = \frac{\alpha^{2} \mu_{I}^{2} - m_{\pi}^{2}}
      {\alpha^{2} \mu_{I}^{2} + 3 m_{\pi}^{2}},
\end{equation}
and
\begin{equation}
  \label{eq:5dspeedofsound}
  c_{s}^{2} = \frac{\alpha^{2} \mu_{I}^{4} - m_{\pi}^{4}}
      {\alpha^{2} \mu_{I}^{4} + 3 m_{\pi}^{4}}.
\end{equation}

\begin{figure}
  \centering
  \scalebox{.5}{\includegraphics{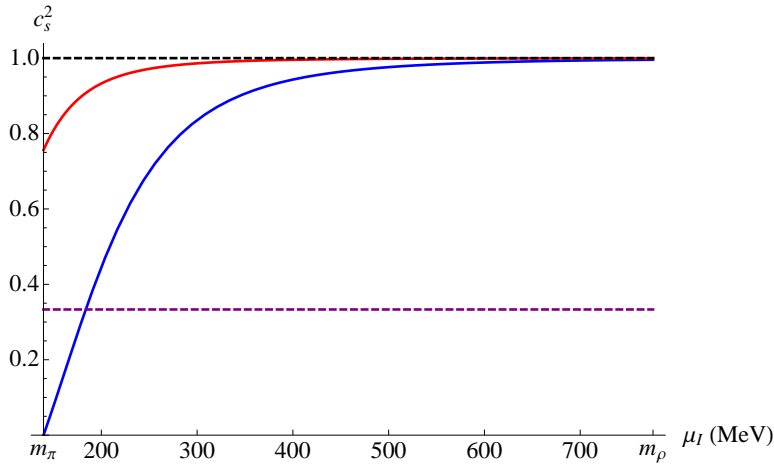}}
  \caption{Speed of Sound.  The upper red curve is the perturbative prediction of the
hard-wall model for the
    speed of sound with \(m_{q} = 2.26\ {\rm MeV}\),
    \(\sigma = (333\ {\rm MeV})^{3}\), and $z_m=1/(323\ {\rm MeV})$.
   This set of parameters gives the value \(\alpha = 3.66\) in
    \eqref{eq:5dspeedofsound}.  The bottom blue curve is the prediction based
on \eqref{eq:n4D}.  The top and bottom dashed lines represent the speed
    of light and the conformal limit $c_s^2=1/3$.}
  \label{fig:speedofsound}
\end{figure}

This is plotted next to the chiral Lagrangian prediction in
Figure \ref{fig:speedofsound}.  
Note that the speed of sound exceeds the sound bound $c_s^2=1/3$ throughout the
pion condensate phase at zero temperature.

\section{A comment on the chiral Lagrangian}
\begin{figure} [ht]
  \centering
  \mbox{\subfigure[\quad \(a_{1}\) less than critical.]{
    \includegraphics[scale=0.4]{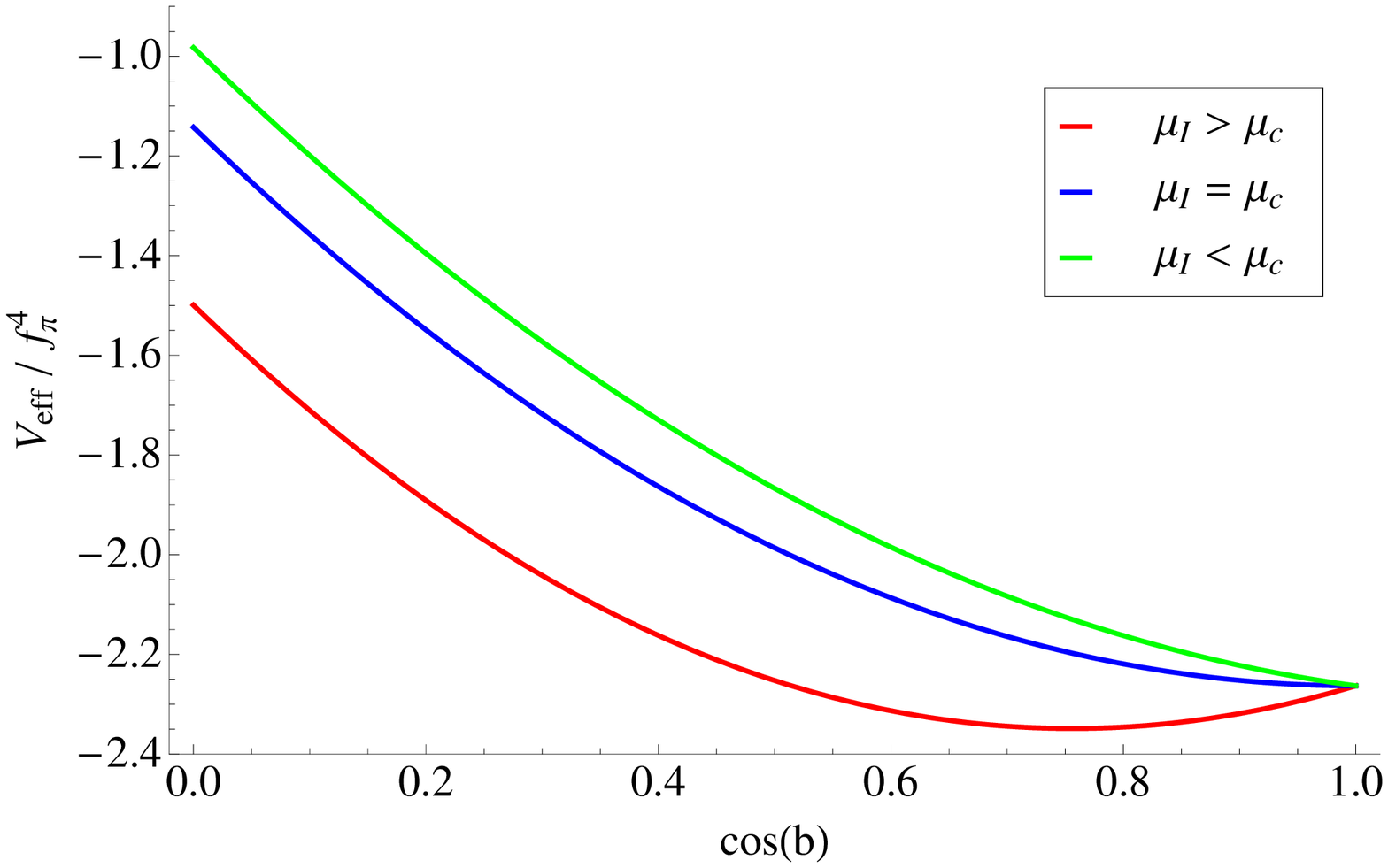}}
    \quad 
    \subfigure[\quad \(a_{1}\) critical.]
      {\includegraphics[scale=0.4]{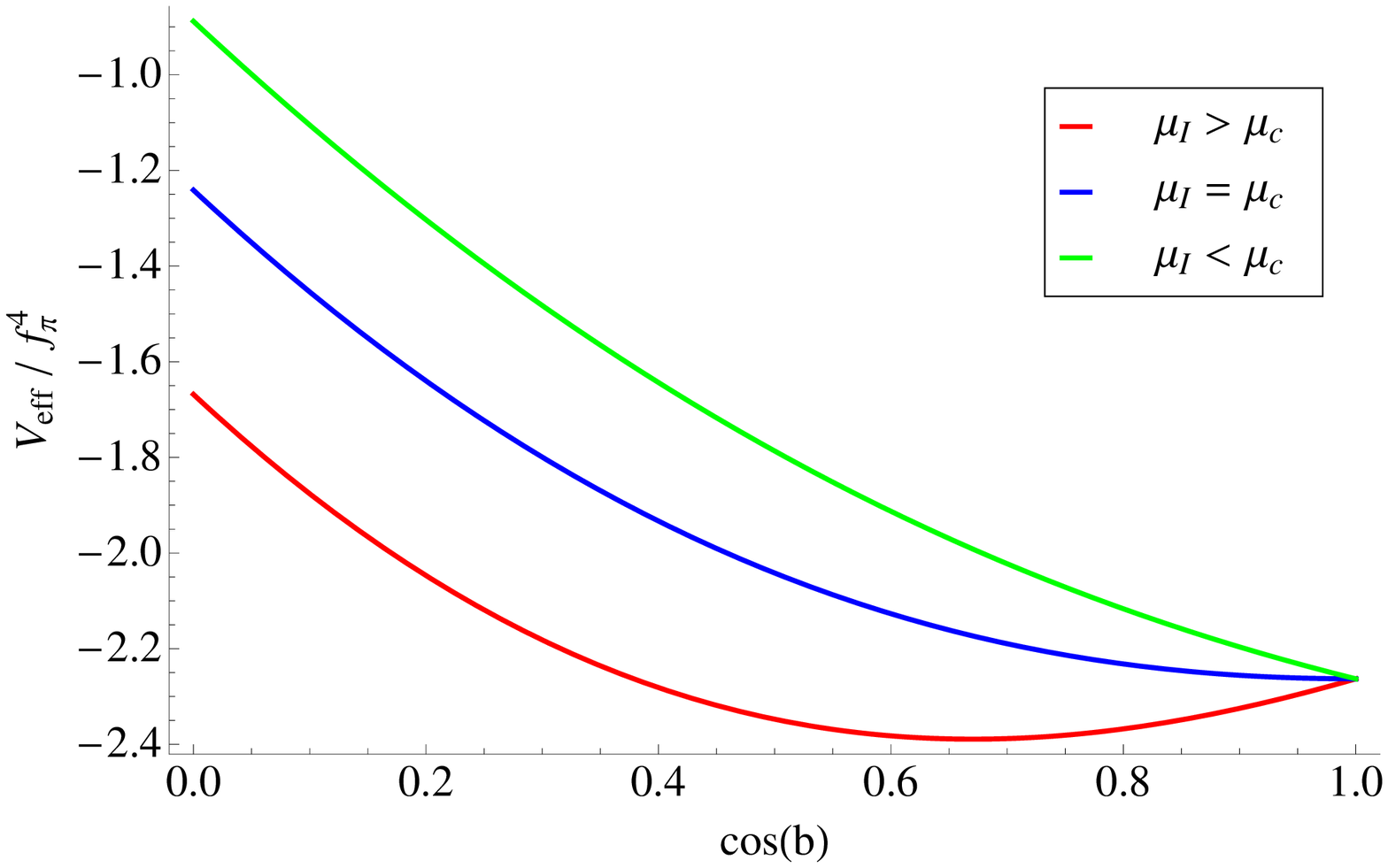}}
  }
  \mbox{
    \subfigure[\quad \(a_{1}\) larger than critical.]
      {\includegraphics[scale=0.4]{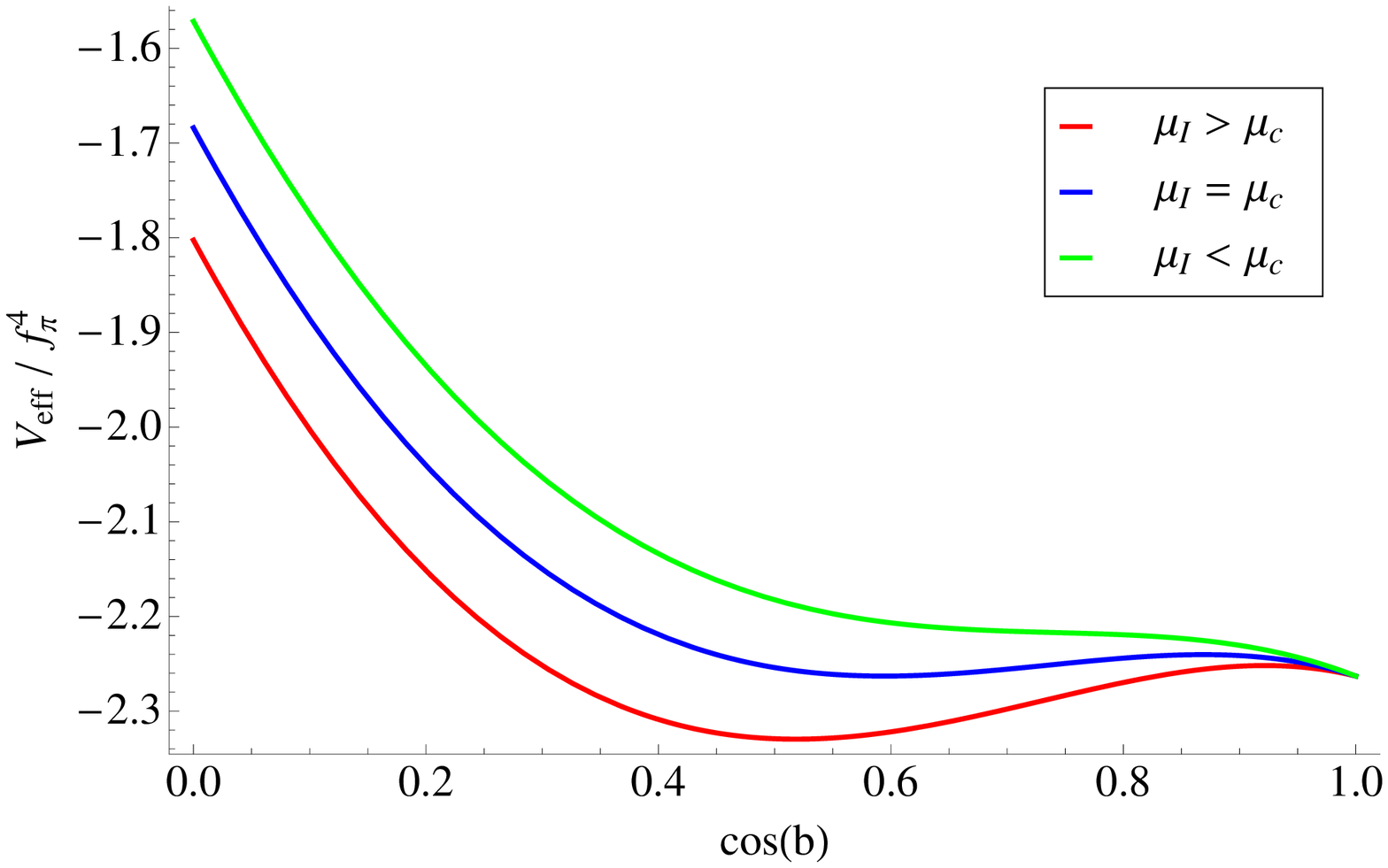}}
  }
  \caption{Each plot shows the phase transition for a different value of the
    \(a_{1}\) parameter.  The critical value of $\mu_I$ for
pion condensation depends on $a_1$.  The three
    curves shown in each plot correspond to \(\mu_{I} < \mu_{c}\), \(\mu_{I} = \mu_{c}\),
    and \(\mu_{I} > \mu_{c}\) (top, middle, and bottom curves, respectively).
    Plot (a) shows the transition for \(a_{1}\) less than the critical
    value.  Plot (b) is with \(a_{1}\) the critical value, while \(a_{1}\) of
    (c) is larger. These plots assumed
    \(m_{\pi} = 139\ {\rm MeV}\) and \(f_{\pi} = 92.4\ {\rm MeV}\).}
  \label{fig:a1values}
\end{figure}

Leading order chiral perturbation theory predicts that the transition to the
pion condensate phase is second order.
We have learned that
gauging the chiral symmetry in the holographic model
qualitatively modifies predictions for pion condensation at 
zero temperature.
The transition becomes first order, and the medium becomes stiff immediately beyond the
phase boundary.    
Including higher derivative terms in the chiral Lagrangian can have
similar consequences, as we will now demonstrate.
Consider the Lagrangian
\begin{equation}
\begin{array} {l}
\mathcal{L} = \frac{f_{\pi}^{2}}{4} \tr\left[D_{\mu} \Sigma D^{\mu}
  \Sigma^{\dagger}\right] + \frac{m_{\pi}^{2} f_{\pi}^{2}}{4} \tr\left[\Sigma
  + \Sigma^{\dagger}\right] + \alpha_{1} \left(\tr\left[D_{\mu} \Sigma
  D^{\mu} \Sigma^{\dagger}\right]\right)^{2}
  \\ \qquad
  + \alpha_{2} \tr\left[D_{\mu} \Sigma
  D_{\nu} \Sigma^{\dagger}\right] \tr\left[D^{\mu} \Sigma
  D^{\nu} \Sigma^{\dagger}\right],
\end{array}
\end{equation}
where for this analysis \(\alpha_{1}\) and \(\alpha_{2}\) are arbitrary parameters.
Once again we take the static
part of the Lagrangian to get an expression for the effective potential.
Defining $\Sigma=\cos b + i\, (n^a \sigma^a) \sin b$, we have
\begin{equation}
V_{eff} (\cos b) = - \frac{\mu_{I}^{2} f_{\pi}^{2}}{2} \left(1 - \cos^{2} b\right)
  \left(1 - n^{3} n^{3}\right)
  - m_{\pi}^{2} f_{\pi}^{2} \cos b - a_{1} \frac{\mu_{I}^{4} f_{\pi}^{2}}{4}
  \left(1 - \cos^{2} b\right)^{2} \left(1 - n^{3} n^{3}\right)^{2},
\end{equation}
where \(a_{1} \equiv \frac{16}{f_{\pi}^{2}} (\alpha_{1} + \alpha_{2})\). 
At the minimum of $V_{eff}$, \(n^{3} = 0\), and  we find a
region of \(a_{1}\) parameter space where the phase transition is first order.
That is, as \(a_{1}\) increases past a critical value 
\(a_{1}^{crit}=1/(2m_\pi^2)\), 
the phase transition changes from second to first order.
This is illustrated in Figure \ref{fig:a1values} .
 However, $f_\pi^2 a_1^{crit}=0.22$ is much larger than the typical
low energy coefficients in the chiral Lagrangian inferred by experiment
 ($l_1(m_\pi)=(-4\pm6)\times10^{-3},\
 l_2(m_\pi)=(9.1\pm0.2)\times10^{-3}$) \cite{Colangelo:2001df,Gasser:1983yg}.

\section{Conclusions}
We have studied pion condensation at zero temperature 
and finite isospin chemical potential in a hard-wall model of
holographic QCD with chiral symmetry breaking and massive pions.  
At the critical point $\mu_I=m_\pi$ the pion condenses, and our perturbative
analysis suggests that the condensate creates a stiff 
medium approaching the Zel'dovich equation of state $p=\varepsilon$.
Sound propagation exceeds the conformal sound bound $c_s^2=1/3$, 
except near the
phase transition boundary if the 5D gauge coupling is small enough.  
The low-energy effective theory for pions as derived from the
hard-wall model indicates that the transition from the hadronic phase 
to the condensate phase 
is first order, except in the limit of vanishing 5D gauge coupling.  
This is in contrast to leading order
chiral perturbation theory, which predicts a second order
transition \cite{Son:2000xc}, and  lattice simulations which also seem to be 
consistent with a second order transition \cite{deForcrand:2007uz}.
We have shown that even in chiral perturbation theory 
the transition can become first order
if higher derivative terms in the chiral 
Lagrangian have large enough coefficients.

\begin{acknowledgments}
We are happy to thank Will Detmold, Brian Smigielski and Andre Walker-Loud
for useful conversations.  This
work was supported by the NSF under Grant PHY-0757481.
\end{acknowledgments}

%\bibliographystyle{/home/dylan/physics/write/bibstyles/hunsrt}
%\bibliography{pionref}
\end{document}